\documentstyle[aps,prl,12pt,epsf]{revtex}
\newcommand{\mevc} {\ifmmode {\rm MeV}/c \else MeV$/c$\fi}
\newcommand{\mevcc} {\ifmmode {\rm MeV}/c^2 \else MeV$/c^2$\fi}
\newcommand{\gevc} {\ifmmode {\rm GeV}/c \else GeV$/c$\fi}
\newcommand{\gevcc} {\ifmmode {\rm GeV}/c^2 \else GeV$/c^2$\fi}
\newcommand{\ra}   {\rightarrow}

\newcommand{\jpsi} {\ifmmode J/\psi \else $J/\psi$\fi}
\newcommand{\vtd}  {\ifmmode |V_{td}| \else $|V_{td}|$\fi}
\newcommand{\vtb}  {\ifmmode |V_{tb}| \else $|V_{tb}|$\fi}
\newcommand{\vts}  {\ifmmode |V_{ts}| \else $|V_{ts}|$\fi}
\newcommand{\vcb}  {\ifmmode |V_{cb}| \else $|V_{cb}|$\fi}

\newcommand{\xs} {\ifmmode x_{\mbox{\sl s}}
                       \else $x_{\mbox{\sl s}}$\fi}
\newcommand{\xd} {\ifmmode x_d \else $x_d$\fi}
\newcommand{\bgam} {\ifmmode \beta\gamma
                       \else $\beta\gamma$\fi}
\newcommand{\Lxy} {\ifmmode L_{\rm xy} \else $L_{\rm xy}$\fi}
\newcommand{\ctau} {\ifmmode c\tau \else $c\tau$\fi}
\newcommand{\Pt} {\ifmmode p_{\rm T} \else $p_{\rm T}$\fi}
\newcommand{\ed} {\ifmmode \varepsilon {\cal D}^2 
		    \else $\varepsilon {\cal D}^2$\fi}
\newcommand{\ptrel} {\ifmmode p_{\rm T}^{\rm rel} 
                       \else $p_{\rm T}^{\rm rel}$\fi}

\newcommand{\As}  {\ifmmode {\cal A} \else ${\cal A}$\fi}
\newcommand{\Dil}{\ifmmode {\cal D} \else ${\cal D}$\fi}
\newcommand{\Do} {\ifmmode {\cal D}_0 \else ${\cal D}_0$\fi}
\newcommand{\Dx} {\ifmmode {\cal D}_+ \else ${\cal D}_+$\fi}
\newcommand{\dmd}{\ifmmode \Delta m_d \else $\Delta m_d$\fi}

\newcommand{\Bds} {\ifmmode B^{**} \else $B^{**}$\fi}
\newcommand{\Dds} {\ifmmode D^{**} \else $D^{**}$\fi}
\newcommand{\pids}{\ifmmode \pi_{**} \else $\pi_{**}$\fi}

\begin{document}

{\bf\Large
Measurement of the $B^0 \bar{B}^0$ oscillation frequency 
using $\pi$-$B$ meson charge-flavor correlations
in $p\bar p$ collisions at $\sqrt{s} = 1.8 \mbox{ TeV}$
}

\font\eightit=cmti8
\def\r#1{\ignorespaces $^{#1}$}
\hfilneg
\begin{sloppypar}
\noindent
F.~Abe,\r {17} H.~Akimoto,\r {39}
A.~Akopian,\r {31} M.~G.~Albrow,\r 7 A.~Amadon,\r 5 S.~R.~Amendolia,\r {27} 
D.~Amidei,\r {20} J.~Antos,\r {33} S.~Aota,\r {37}
G.~Apollinari,\r {31}
% ...
\end{sloppypar}
\vskip .026in
\begin{center}
(CDF Collaboration)
\end{center}

\vskip .026in
\begin{center}
\r 1  {\eightit Argonne National Laboratory, Argonne, Illinois 60439} \\
\r 2  {\eightit Istituto Nazionale di Fisica Nucleare, University of Bologna,
I-40127 Bologna, Italy} \\
\r 3  {\eightit Brandeis University, Waltham, Massachusetts 02254} \\
\r 4  {\eightit University of California at Los Angeles, Los 
Angeles, California  90024} \\  
\r 5  {\eightit University of Chicago, Chicago, Illinois 60637} \\
\r 6  {\eightit Duke University, Durham, North Carolina  27708} \\
\r 7  {\eightit Fermi National Accelerator Laboratory, Batavia, Illinois 
60510} \\
\r 8  {\eightit University of Florida, Gainesville, FL  32611} \\
\r 9  {\eightit Laboratori Nazionali di Frascati, Istituto Nazionale di Fisica
               Nucleare, I-00044 Frascati, Italy} \\
\r {10} {\eightit University of Geneva, CH-1211 Geneva 4, Switzerland} \\
\r {11} {\eightit Harvard University, Cambridge, Massachusetts 02138} \\
\r {12} {\eightit Hiroshima University, Higashi-Hiroshima 724, Japan} \\
\r {13} {\eightit University of Illinois, Urbana, Illinois 61801} \\
\r {14} {\eightit Institute of Particle Physics, McGill University, Montreal 
H3A 2T8, and University of Toronto,\\ Toronto M5S 1A7, Canada} \\
\r {15} {\eightit The Johns Hopkins University, Baltimore, Maryland 21218} \\
\r {16} {\eightit Institut f\"{u}r Experimentelle Kernphysik, 
Universit\"{a}t Karlsruhe, 76128 Karlsruhe, Germany} \\
\r {17} {\eightit National Laboratory for High Energy Physics (KEK), Tsukuba, 
Ibaraki 305, Japan} \\
\r {18} {\eightit Ernest Orlando Lawrence Berkeley National Laboratory, 
Berkeley, California 94720} \\
\r {19} {\eightit Massachusetts Institute of Technology, Cambridge,
Massachusetts  02139} \\   
\r {20} {\eightit University of Michigan, Ann Arbor, Michigan 48109} \\
\r {21} {\eightit Michigan State University, East Lansing, Michigan  48824} \\
\r {22} {\eightit University of New Mexico, Albuquerque, New Mexico 87131} \\
\r {23} {\eightit The Ohio State University, Columbus, OH 43210} \\
\r {24} {\eightit Osaka City University, Osaka 588, Japan} \\
\r {25} {\eightit Universita di Padova, Istituto Nazionale di Fisica 
          Nucleare, Sezione di Padova, I-36132 Padova, Italy} \\
\r {26} {\eightit University of Pennsylvania, Philadelphia, 
        Pennsylvania 19104} \\   
\r {27} {\eightit Istituto Nazionale di Fisica Nucleare, University and Scuola
               Normale Superiore of Pisa, I-56100 Pisa, Italy} \\
\r {28} {\eightit University of Pittsburgh, Pittsburgh, Pennsylvania 15260} \\
\r {29} {\eightit Purdue University, West Lafayette, Indiana 47907} \\
\r {30} {\eightit University of Rochester, Rochester, New York 14627} \\
\r {31} {\eightit Rockefeller University, New York, New York 10021} \\
\r {32} {\eightit Rutgers University, Piscataway, New Jersey 08855} \\
\r {33} {\eightit Academia Sinica, Taipei, Taiwan 11530, Republic of China} \\
\r {34} {\eightit Texas A\&M University, College Station, Texas 77843} \\
\r {35} {\eightit Texas Tech University, Lubbock, Texas 79409} \\
\r {36} {\eightit Istituto Nazionale di Fisica Nucleare, University of Trieste/
Udine, Italy} \\
\r {37} {\eightit University of Tsukuba, Tsukuba, Ibaraki 315, Japan} \\
\r {38} {\eightit Tufts University, Medford, Massachusetts 02155} \\
\r {39} {\eightit Waseda University, Tokyo 169, Japan} \\
\r {40} {\eightit University of Wisconsin, Madison, Wisconsin 53706} \\
\r {41} {\eightit Yale University, New Haven, Connecticut 06520} \\
\end{center}

\begin{abstract}
We present a measurement of the $B^0 \leftrightarrow \bar{B}^0$ 
oscillation frequency
using a flavor tagging method based on
correlations of $B$ meson flavor with the charge of other particles 
produced
in  $p\bar p$ collisions at $\sqrt{s} = 1.8 \mbox{ TeV}$. 
Such correlations are expected to arise from 
$b$~quark hadronization and from $B^{**}$ decays.
We partially reconstruct $B$ mesons using the 
semileptonic decays $B^0 \rightarrow \ell^+ D^{(*)-} X$
and $B^+ \rightarrow \ell^+ \bar D^0 X$.
From the oscillation frequency, we obtain the 
mass difference between the two $B^0$ mass eigenstates, 
$\Delta m_d = 0.471   
	^{+0.078}_{-0.068} 
	\mbox{(stat)} 
	\pm 0.034     
	\mbox{(syst)} \>\hbar\mbox{ ps}^{-1}$,
and measure the efficiency and purity of this flavor tagging
method for both charged and neutral $B$ mesons.
\end{abstract}

\vspace{.3cm}

\centerline{PACS numbers: 13.20.He, 14.40.Nd}
\vspace{.3cm}

The $B^0$ meson is a bound state of a $\bar{b}$ quark and a $d$ quark.
Second order electroweak processes in which
$\bar{b} \to \bar{d}$ while
$d \to b$
result in the transition of
$B^0$ mesons into $\bar{B}^0$ mesons and thus
$B^0 \leftrightarrow \bar{B}^0$ oscillations.
The $b \to d$ coupling occurs via a virtual top quark, thus
the frequency of these oscillations
is sensitive to the magnitude of the element $V_{td}$ of 
the Cabbibo-Kobayashi-Maskawa matrix~\cite{ref:CKM}. 
For an initially pure sample of $B^0$ mesons 
(at $t=0$), the numbers of $B^0$ ($N_+$) and $\bar B^0$ ($N_-$)
mesons at proper time $t$ are given by
\begin{equation}
  N_{\pm}(t) = N_+(0)\frac{e^{-t/\tau}}{2}(1\pm\cos\dmd t),
	\label{eq:mixing}
\end{equation}
where $\tau$ is the lifetime of the $B^0$ meson
and $\Delta m_d$ is the mass difference between the mass eigenstates of 
the $B^0$-$\bar{B}^0$ system.
In this paper we determine $\Delta m_d$ by measuring the frequency of
$B^0 \leftrightarrow\bar{B}^0$ flavor oscillations 
using partially reconstructed
semileptonic decays of the $B$~meson to $\ell D^{(*)} X$.
The data used in this analysis were collected with the Collider Detector
at Fermilab (CDF), at the Tevatron $p\bar{p}$ Collider at $\sqrt{s}=1.8$ TeV,
and correspond to an integrated luminosity of $\sim 110$~pb$^{-1}$.

To extract
\dmd\ the proper time of the $B$ decay is required, as well as 
the flavor of the
$B$ meson at the times of its decay and production.
While the $B$ flavor at decay is determined by its 
decay products, 
the determination of the initial $B$ flavor is experimentally 
challenging. 
Techniques to determine the initial $B$ flavor in
several previous measurements of 
\dmd~\cite{ref:mixing} relied on identifying the flavor from the
other $b$ hadron in the event ({\em e.g.}, using the lepton charge from the
semileptonic decay of this other hadron),
and are thus referred to as Opposite Side Tagging (OST).

It has been suggested~\cite{ref:Rosner} that the electric charge of
particles produced near a $B$ meson could also be used to
determine its initial flavor, providing a basis for Same Side Tagging (SST).
For example, if a $\bar b$ quark combines with a $u$~quark to form a
$B^+$ meson 
the remaining $\bar u$ quark may combine with a
$d$~quark to form a $\pi^-$
(Throughout this paper, reference to a specific
particle state implies the charge conjugate state as well).
Similarly, if a $\bar b$ quark hadronizes to form a $B^0$ meson, the
associated pion would be a $\pi^+$.
Another source of correlated pions are decays of the
orbitally excited ($L=1$) $B$ mesons ($B^{**}$)~\cite{ref:LEP},
 {\em i.e.},
$B^{**0} \ra B^{(*)+} \pi^-$ or
$B^{**+} \ra B^{(*)0} \pi^+$.
In a hadron collider experiment with central rapidity coverage
such as CDF, SST methods are attractive since they are expected to have 
significantly higher efficiency than the OST methods.  These methods
can also be used for measurements of CP violation~\cite{ref:TDR}.

In this paper we extract \dmd\ by applying an SST method 
to a sample of events containing a lepton and
a reconstructed $D$ meson from $B$ decay.
To determine the initial flavor of the $B$ meson, we select
one charged track that we will generically refer to as a ``pion'',
and use its charge as a tag.  We do not attempt to distinguish the 
hadronization pions from those originating from $B^{**}$ decays. 
The lepton charge tags the $B$-flavor at decay time.
We classify the $B$-$\pi$ combinations 
as right-sign (RS: $B^+\pi^-$ and $B^0\pi^+$) 
or wrong-sign (WS: $B^+\pi^+$ and $B^0\pi^-$).
We form the asymmetry in the RS and WS combinations,
${\cal A}(ct)\equiv (N_{RS}(ct)-N_{WS}(ct))/( N_{RS}(ct)+N_{WS}(ct))$,
as a function of the proper decay length $ct$.
For $B^+$ mesons, we expect an asymmetry independent of $ct$:
${\cal A}^+(ct) = \mbox{constant} \equiv {\cal D}_+$.
The quantity ${\cal D}_+$ is called the dilution and it is
a direct measure of the SST purity, {\em i.e.}, $(1+{\cal D})/2$ is the
fraction of correctly tagged events.
Due to $B^0\bar{B}^0$ mixing, ${\cal A}(ct)$ for the neutral
$B$ mesons will vary as a function of the proper decay length $ct$.
From eq.~(\ref{eq:mixing}) and the definition of the $B^0$ asymmetry,
it follows that the latter is expected to oscillate as
${\cal A}^0(ct) = {\cal D}_0 \cdot\cos(\dmd t)$.
Mistags, {\em i.e.}, incorrect flavor determinations, result in
a decrease of the oscillation amplitude by the $B^0$ dilution
factor ${\cal D}_0$.
We measure the asymmetry as a function of the
proper decay length $ct$
for both $B^+$ and $B^0$ mesons,
and fit them with their expected time dependence, obtaining
$\dmd$, ${\cal D}_0$ and ${\cal D}_+$.
Ref.~\cite{ref:dif_dil} discusses several effects that can result
in $\Do \neq \Dx$, and in this analysis we impose
no constraints on their relative values.

The CDF detector is discussed in detail elsewhere~\cite{ref:cdf_detect},
and only the features most relevant to this 
analysis are described here.  
In CDF, the positive $z$ axis is pointed in the proton direction,
$\theta$ is the polar angle and 
$\phi$ is the azimuthal angle.  The pseudorapidity, $\eta$, is defined as 
$-\ln\left[\tan(\theta/2)\right]$.  
We use two devices inside a $1.4$~T solenoid to measure tracks from
charged particles:
the central tracking chamber (CTC) and the silicon vertex detector (SVX).
The combined CTC+SVX tracking system 
covers the pseudorapidity interval $|\eta| < 1.1$,
and gives a resolution on the transverse momentum
with respect to the beam axis, $\Pt$, of
$\delta(\Pt)/\Pt = ((0.0066)^2 + (0.0009 \Pt)^2)^{1/2}$
and a resolution on the track impact parameter,
$d_0$, defined as the distance of closest
approach to the beamline,
of about $(13 + 40/\Pt)~\mu$m, where $\Pt$ is in GeV/$c$. 
Electromagnetic (CEM) and hadronic (CHA) calorimeters are located
outside the solenoid and are surrounded by
the central muon chambers (CMU) followed by
the central upgrade muon chambers (CMP).

The data were recorded using an inclusive lepton ($e$ and $\mu$) trigger.
The $E_T$ threshold for the single electron trigger was 8 GeV,
where $E_T \equiv E\sin\theta$, and $E$ is the energy measured in the CEM.
The single muon trigger required a charged track with $\Pt > 7.5$ GeV/$c$ 
in the CTC with matched track segments in both the CMU and CMP systems. 
Details of the identification of electrons and muons are described in 
references~\cite{ref:electron} and~\cite{ref:cdf_life}. 

\begin{figure}[p]\centering
\centerline{
\epsfysize=18cm
\epsffile{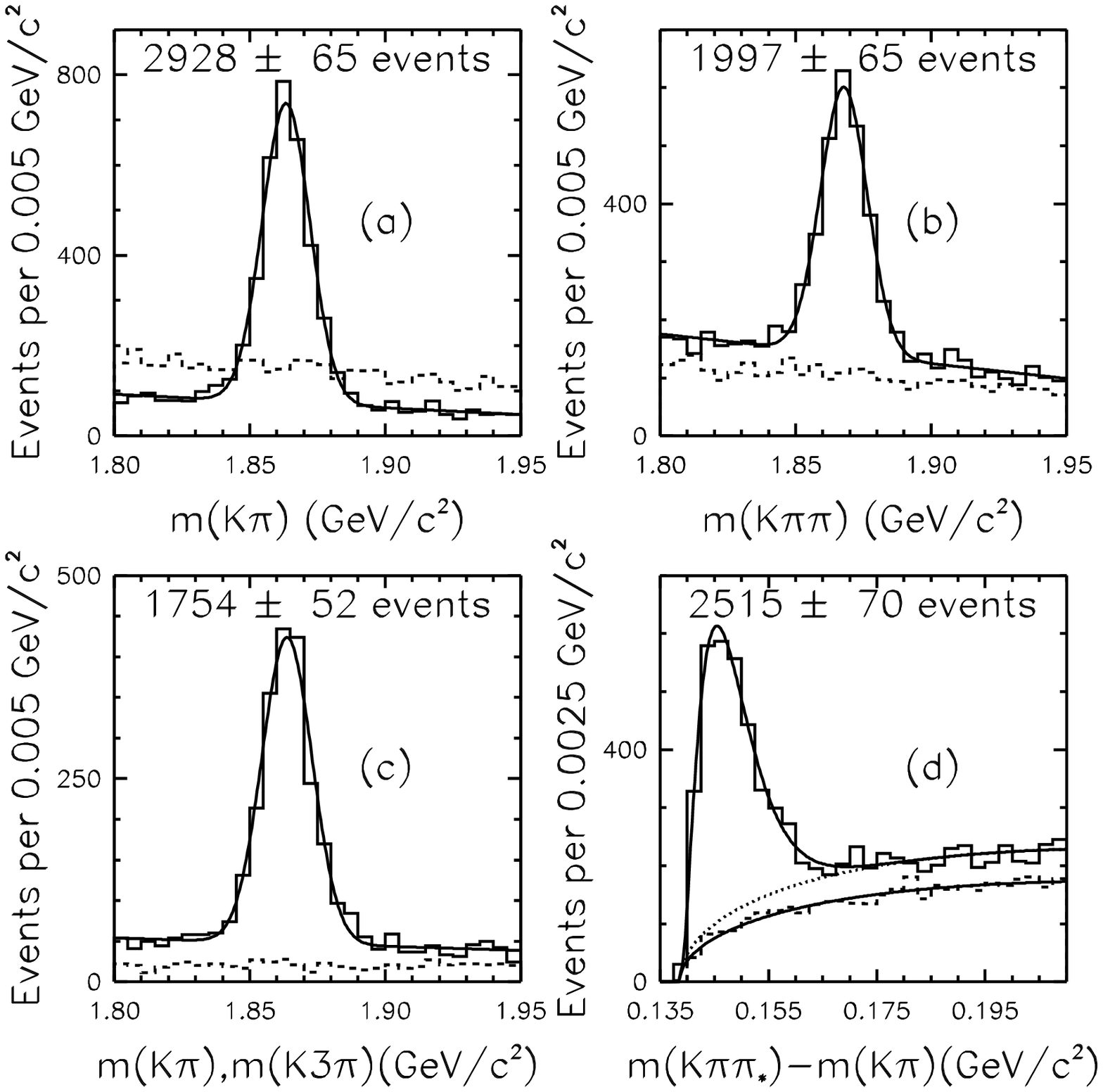} }
\caption{Invariant mass distributions of the five $B \to \ell^+ D^{(*)}X$ 
	signatures used in the analysis.
	The solid histograms correspond to the ``right'' lepton-kaon charge
	correlation ($\ell^+K^+$), 
	and the dashed histograms the ``wrong'' one ($\ell^+K^-$).
	The solid lines are the fits to the $\ell^+K^+$ distributions,
	and the resulting number of signal events is given on each figure.
	(a) $K^+\pi^-$ mass in $\ell^+\bar{D}^0 X$
		($\bar D^0\to K^+\pi^-$).
	(b) $K^+\pi^-\pi^-$ mass in $\ell^+ D^{-}X$
		($D^- \to K^+\pi^-\pi^-$).
	(c) $\bar{D}^0$ candidate mass in $\ell^+ D^{*-}X$
		($D^{*-} \to \bar D^0 \pi_*^-$,
		$\bar D^0\to K^+\pi^-$ and  $\bar D^0\to K^+\pi^-\pi^+\pi^-$).
	(d) $K^+\pi^-\pi_*^- - K^+\pi^-$ 
		mass difference in $\ell^+ D^{(*)-}X$
		($D^{*-} \to \bar D^0 \pi_*^-$,
		$\bar D^0\to K^+\pi^-\pi^0$).
	The $\ell^+K^+$ background shape (dotted line)
	was determined from the fit (lower solid curve) of the $\ell^+K^-$ 
	distribution.
	\label{mass4.pretty}
	}
\end{figure}

We use the decay chains
$B^0 \rightarrow \nu \ell^+ D^{(*)-}$, with $ D^-\rightarrow K^+\pi^-\pi^-$,
or $D^{*-} \rightarrow \bar D^0 \pi_*^-$
followed by $\bar D^0$ decaying to $K^+\pi^-$, $K^+\pi^-\pi^+\pi^-$,
or $K^+\pi^-\pi^0$, where $\pi_*^-$ denotes
the low-momentum (soft) pion from the $D^{*-}$ decay.
We use
$B^+ \rightarrow \nu \ell^+ \bar D^0$, with
$\bar D^0\rightarrow K^+\pi^-$, where the $\bar D^0$ is required not to form
a $D^{*-}$ candidate with another $\pi$ candidate in the event.
We reconstruct the $D$ meson candidates using tracks in a circle 
of unit radius in $\eta$-$\phi$ space  around the lepton.
For details, see~\cite{ref:FumiLifetime}. 
We use only tracks that include SVX information.
The decay products from the $D$ mesons are required to 
be significantly displaced from the interaction point (primary vertex) 
of the event.
The mass distributions of the four decay signatures with fully reconstructed
$D$~mesons are shown in Fig.~\ref{mass4.pretty}a, b and c, while 
the distribution of $\Delta m = m(K\pi\pi_*)-m(K\pi)$
for the signature with $D^{*-} \rightarrow \bar D^0 \pi_*^-$,
followed by $\bar D^0 \rightarrow K^+\pi^-\pi^0$ 
(the $\pi^0$ is not reconstructed)
is displayed in Fig.~\ref{mass4.pretty}d.

To select the SST pion, we consider
all tracks that are within the circle in $\eta$-$\phi$ 
space
of radius $0.7$ centered
around the direction of the $B$ meson, approximated by 
$\vec{p}(\ell)+\vec{p}(D)$.
SST candidate tracks should originate from
the $B$ production point (the primary vertex of the event),
and are therefore required 
to satisfy $d_0/\sigma_{d_0}<3$,
where $\sigma_{d_0}$ is the uncertainty on $d_0$. 
String fragmentation models~\cite{ref:string_models} 
indicate that particles produced in the $b$-quark hadronization chain 
have low momenta transverse to the direction of the $b$-quark momentum.
We thus select as the tag the track that has the minimum
component of momentum, $\ptrel$, orthogonal to the momentum sum of
the track and the $B$ meson.  
We define the tagging efficiency, $\epsilon$, as 
the fraction of $B$ candidates with at least one track satisfying
the above requirements.  We measure $\epsilon \approx 70\%$,
independent of the decay signature used.
On average, there are about $2.2$ SST candidate tracks per 
$B$ candidate.

For each of the five decay signatures, we subdivide the candidates into
six bins in proper decay length, $ct$.  
The measurement of $ct$ 
begins with finding the $D$ decay vertex.
The $D$ candidate trajectory is intersected with the lepton track 
(and the momentum of $\pi_*$ from the $D^*$, if present) 
to form the $B$ decay vertex.
We define  $\Lxy^{B}$ to be, in the plane transverse to the
beam axis, the distance between the primary and the $B$ vertex
projected along the $B$-momentum vector.
We estimate $\Pt^B$
from the $p_T$ of the visible decay products,
$\Pt^{\ell D}=|\vec{p}_{\rm T}(\ell)+\vec{p}_{\rm T}(D^{(*)})|$, and the mean
of the distribution of the momentum ratio 
${\cal K} \equiv \langle\Pt^{\ell D}/\Pt^B\rangle$ obtained
from a Monte Carlo simulation~\cite{ref:bgen}.  
The $\Pt^{\ell D}/\Pt^B$ distribution has
a mean of $\sim 85\%$ and RMS of $\sim 12\%$.
The proper decay length of the 
$B$ meson is then given by 
$ct = L^{B}_{xy}(m_{B}/\Pt^{\ell D}){\cal K}$~\cite{ref:thesis}.

\begin{figure}[tbp]
\centerline{
\epsfysize=18cm
\epsffile{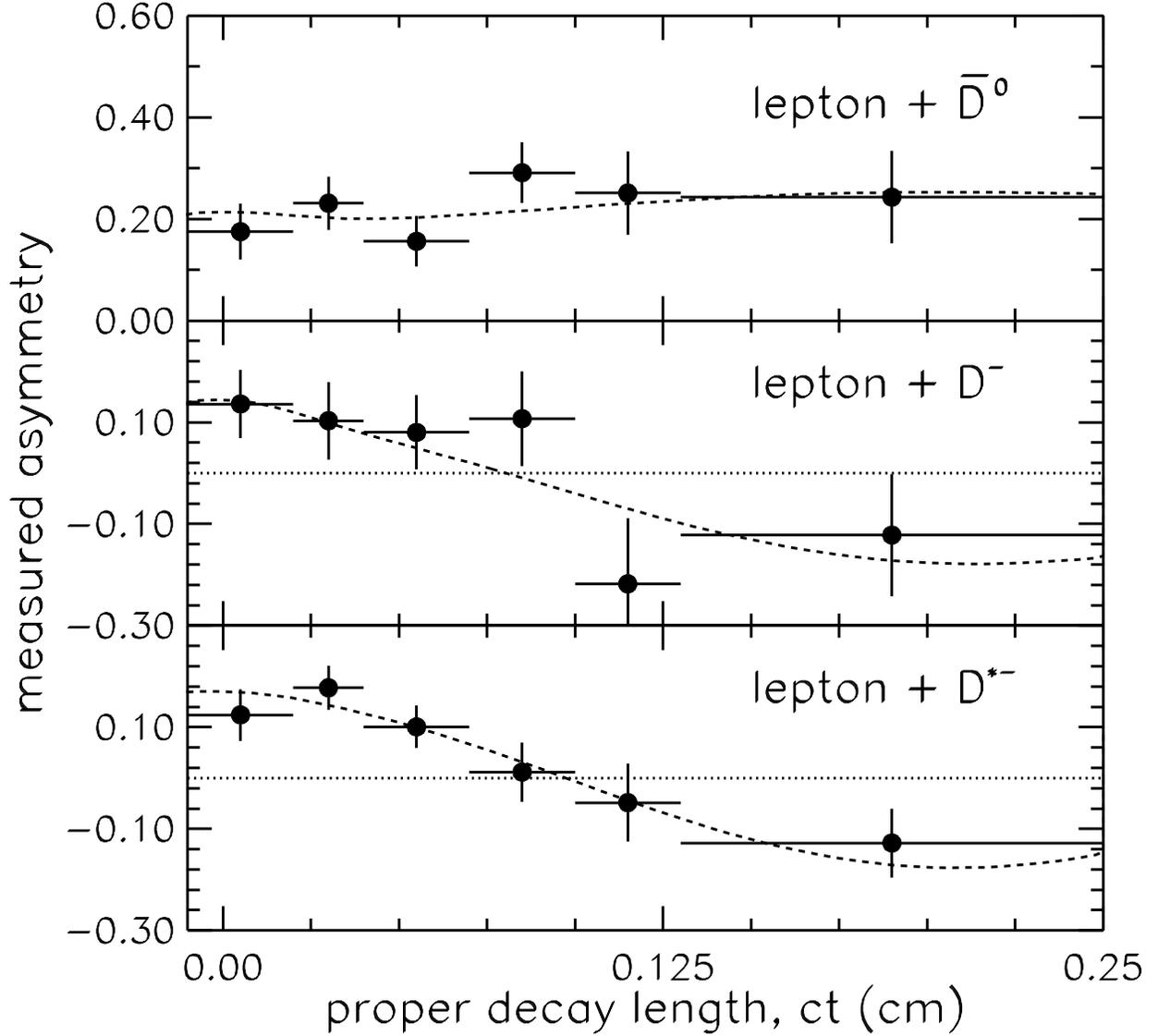}
}
\caption{Measured asymmetries as a function of the proper decay length, $ct$,
	for the decay signatures: $\ell^+ \bar{D}^0$
	(dominated by $B^+$), 
	$\ell^+ D^-$ and the sum of all three $\ell^+ D^{*-}$
	(dominated by $B^0$). We fit the three $\ell^+ D^{*-}$ signature
	separately,  but combine them for display purposes.
	The dashed line is the result of the fit.
	\label{fig:result}
	}
\end{figure}

We measure the asymmetry ${\cal A}^{(m)}(ct)$, 
in each $ct$ bin by simultaneously fitting 
the mass distributions for candidates tagged with a 
RS or WS pion.  The measured asymmetries are shown in Fig.~\ref{fig:result}.
If the $\ell^+\bar D^0$ and $\ell^+D^{(*)-}$ signatures
were pure signals of $B^+$ and $B^0$ decays, we could simply
extract \dmd\ using the time-dependence
of ${\cal A}^0(ct)$. However,
the signatures are mixtures of $B^+$ and $B^0$ decays,
and thus ${\cal A}^{(m)}(ct)$ is a linear combination of the
true asymmetries ${\cal A}^0(ct)$ and ${\cal A}^+(ct)$.
To extract $\Delta m_d$, ${\cal D}_0$ and ${\cal D}_+$,
it is necessary to determine 
the {\em sample composition} of each $\ell^+ D^{(*)}$ signature,
by which we mean
the fractions of the $\ell^+ D^{(*)}$ 
candidates originating from the decays of the $B^0$ and $B^+$ mesons.
Because a $B^+$ is associated with a $\pi^-$, whereas 
an unmixed $B^0$ is associated with a $\pi^+$, the observed
asymmetries are reduced by cross-contamination. 
To obtain the
true asymmetries ${\cal A}^0$ and ${\cal A}^+$,
we introduce sample composition parameters
($f^{**}$, $P_V$, $\xi_{norm}$, $R_f$,  $\epsilon(\pi_*)$, and
 $\tau_{B^+}/\tau_{B^0}$, defined below) to describe this cross-contamination
and vary these parameters
in a simultaneous fit to all of the observed asymmetries.

Cross-contamination can arise
if the soft pion $\pi_*^-$ from the $D^{*-}$ decay is not
identified -- the decay sequence $B^0 \rightarrow \nu\ell^+ D^{*-}$
will be reconstructed as $\ell^+ \bar{D}^0$, that is, as a $B^+$ candidate.
We quantify this effect by the efficiency for reconstructing
the soft pion, $\epsilon(\pi_*)$.  
We determine $\epsilon(\pi_*)$ by comparing the 
fraction, $R^*$, of all $\ell^+ \bar{D}^0 X$
candidates for which we find a $\pi_*^-$ candidate,
with the prediction from the other sample composition parameters.
(The final result is $\epsilon(\pi_*) = 0.85 \pm  0.07$).
Since $\bar{D}^{*0}$ does not decay to $D^- \pi^+$, there is no
cross-contamination from $B^+ \rightarrow \nu\ell^+ \bar{D}^{*0}$
into the $B^0$ sample.

Another source of cross-contamination arises from semileptonic
$B$ decays involving $P$-wave $D^{**}$ resonances
as well as non-resonant $D^{(*)}\pi$ pairs,
which cannot be easily recognized and removed from the sample.
For example, the decay sequence $B^0 \rightarrow \nu\ell^+ D^{**-}$,
followed by $D^{**-} \rightarrow \bar{D}^0 \pids^-$ (where by $\pids$ we 
denote the pion originating from $D^{**}$ decay)
will be reconstructed as $\ell^+\bar{D}^0$, because of the
missed $\pi_{**}^-$; again, a $B^0$ decay is
misclassified as a $B^+$ candidate.
We quantify this effect by the
parameter $f^{**}$, which is the ratio of the branching fraction
${\cal B}(B\rightarrow\nu\ell \Dds)$ to the inclusive semileptonic 
$B$ branching fraction, ${\cal B}(B\rightarrow\ell\nu X)$.  
We use $f^{**} = 0.36 \pm 0.12$~\cite{cleo**}.
The fractions of $B^+$ and $B^0$ decays in each decay signature
are also affected by the relative abundance of the four 
possible spin-parity $D^{**}$ states, 
some of which decay only to $D^*\pi$ and others to $D\pi$.
We define
$P_V = {\cal B}(B\to D^{**} \to D^*\pi)/\left[
 {\cal B}(B\to D^{**} \to D^*\pi) +  {\cal B}(B\to D^{**} \to D\pi)\right]$,
which we leave as a free parameter in our fit.

The sample composition also depends on $f$ and $f^*$, the ratios
of the branching fractions $B \to \nu\ell D$ and 
$B \to \nu\ell D^*$ to the inclusive semileptonic $B$ branching fraction.  
We define $R_f \equiv f^*/f$, use $R_f = 2.5 \pm 0.6$~\cite{PDG}, and assume
$f + f^* + f^{**} = 1$.
The lifetime ratio $\tau_{B^+}/\tau_{B^0}$ is another sample composition 
parameter: $\tau_{B^+} \neq \tau_{B^0}$ implies 
${\cal B}(B^+\rightarrow\ell^+\nu X) \neq {\cal B}(B^0\rightarrow\ell^+\nu X)$,
as well as a $ct$-dependent sample composition.
We use $\tau_{B^+}/\tau_{B^0} = 1.02 \pm 0.05$~\cite{PDG}.

The tagging is further complicated
when a $\pi_{**}^\pm$ from $D^{**}$ decay is present. The $\pi_{**}^\pm$
may be incorrectly selected as the SST pion,
always resulting in a RS correlation.
The requirement $d_0/\sigma_{d_0}<3$ reduces this effect: the \pids\ 
originates from the $B$~meson {\em decay} point, whereas the appropriate
tagging track originates from the $B$~meson {\em production} point.  
The contamination of the tagging pions from $\pids^\pm$
decreases with increasing $ct$.  It
is quantified by the parameter $\xi$, defined as the probability
of selecting the \pids\ as the tag in a tagged event in which a $\pids^\pm$ 
was produced.  We use a Monte Carlo simulation~\cite{ref:pythia}
of $b$ quark fragmentation and decay 
to determine $\xi$ as a function of proper decay time.  
We then use our data to determine a normalization factor for $\xi$,
$\xi_{norm}$, from the fraction, $R^{**}$,
of all tagged events in which a $\pi_{**}$ was selected as the tag.
We measure $R^{**}$ from the
distribution of impact parameter, $d_B$, of the SST pion with 
respect to the $B$ decay point.
In this study, the $d_0/\sigma_0 < 3$ requirement was removed 
to increase the number of events.
We fit the RS distribution of 
$d_B/\sigma_{d_B}$ with the sum of the
scaled WS distribution and a Gaussian of unit width centered at zero.  
The area of this Gaussian yields the total number of SST pions originating 
from $D^{**}$ decays.  
We extract $\xi_{norm}$ by comparing
$R^{**}$ with the prediction from the other
sample composition parameters.
We measure separate values of $R^{**}$ for $\ell D$ and $\ell D^*$
signatures, and their ratio constrains $P_V$.
($\xi_{norm} = 0.8 \pm  0.2$ and $P_V = 0.3 \pm 0.3$ are the final results).

The mass difference \dmd\ and 
the dilutions ${\cal D}_0$ and ${\cal D}_+$ are determined
from a $\chi^2$ fit to 
the measured asymmetries ${\cal A}^{(m)}$.
In the fit, $f^{**}$, $R_f$ and $\tau_{B^+}/\tau_{B^0}$ vary within
their uncertainties, 
while $\epsilon(\pi_*)$, $P_V$, and $\xi_{norm}$ vary within
constraints imposed through the measured values of $R^*$ and $R^{**}$.
The prediction for an observed $B^0$ asymmetry is
\begin{equation}
\bar{\cal A}_k 
   =   \alpha^{0}   _k     {\cal A}^0
   +   \alpha^{+}   _k (-{  \cal A}^+)
   +   \alpha^{**}_k (+1)
\label{eq:A(k,ct)}
\end{equation}
where all information about the sample composition is contained in 
the coefficients $\alpha_k^{0,+}$~\cite{ref:thesis}.
The second term in~(\ref{eq:A(k,ct)}) describes the cross-contamination
occurring with the minus sign due to the opposite charge correlation
for $B^+$, while the third term
describes the effect of selecting the $\pids^\pm$ as a tag.

The result of the fit, overlaid onto the measured asymmetries, 
is displayed in Fig.~\ref{fig:result}.  The oscillation in the
neutral $B$ signatures is clearly present, giving 
$\Delta m_d = 0.471 \>\hbar\mbox{ ps}^{-1}$.
The fit has a $\chi^2$ of $26.5$ for $30$ degrees of freedom.
We assign the uncertainties from the sample composition parameters,
which are dominated by the uncertainty on $f^{**}$,
to the systematic uncertainties on $\Delta m_d$, ${\cal D}_0$ or ${\cal D}_+$.
Other sources of systematic uncertainty are
much smaller, and
are due to uncertainties in the Monte Carlo simulation 
as well as the presence of physics background processes that
can mimic the $\ell^+ D^{(*)}X$ signatures.
The systematic uncertainties are given in 
Table~\ref{tab:systematics}.
The final result for the mixing frequency is 
$\Delta m_d = 0.471   
	^{+0.078}_{-0.068} 
	\mbox{(stat)} 
	\pm 0.034     
	\mbox{(syst)} \>\hbar\mbox{ ps}^{-1}$.
We also obtain the following values for the neutral and charged
meson tagging dilutions: 
${\cal D}_0 = 0.18    
	\pm 0.03      
	\mbox{(stat)}       
	\pm 0.02      
	\mbox{(syst)}$ 
and
${\cal D}_+ = 0.27    
	\pm 0.03      
	\mbox{(stat)}       
	\pm 0.02      
	\mbox{(syst)}$.
The fit indicates that
$\sim 82\%$ of the $\ell^+\bar{D}^0 X$ 
signature comes from $B^+$ decays, while $\sim 80\%$ of the $\ell^+D^- X$  and
$\sim 95\%$ of the $\ell^+ D^{*-} X$ originate from $B^0$.  The $B^0$ component
of the $\ell^+\bar{D}^0 X$ signature can be seen as a small 
anti-oscillation in Fig.~\ref{fig:result}, top.

In conclusion, we have applied a Same Side Tagging technique to 
samples of $B^0 \rightarrow \ell^+ D^{(*)-} X$
and $B^+ \rightarrow \ell^+ \bar D^0 X$ decays
in $p\bar p$ collisions. 
The measurement of the asymmetry between tag-charge and $B$-flavor as
a function of proper time for neutral $B$ mesons
results in the observation of a time-dependent oscillation
$B^0 \leftrightarrow \bar B^0$, with the oscillation frequency
$\Delta m_d = 0.471   
	^{+0.078}_{-0.068} 
	\mbox{(stat)} 
	\pm 0.034     
	\mbox{(syst)} \>\hbar\mbox{ ps}^{-1}$, in good agreement with
the world average 
$\Delta m_d = 0.474 \pm 0.031 \>\hbar\mbox{ ps}^{-1}$~\cite{PDG}.
This 
establishes the effectiveness 
of Same Side Tagging for the first time in hadronic collisions.

We thank the Fermilab staff and the technical staffs of the
participating institutions for their vital contributions. This work was
supported by the U.S. Department of Energy and the National Science 
Foundation; the Italian Istituto Nazionale di Fisica Nucleare; the 
Ministry of Education, Science and Culture of Japan; the Natural 
Sciences and Engineering Research Council of Canada; the National 
Science Council of the Republic of China; and the A.P.~Sloan Foundation.

\begin{table}[htbp]
\begin{tabular}{|l|c|c|c|} 
\hline
Source 	& $\sigma(\Dx)$ & $\sigma(\Do)$ & $\sigma(\dmd) (\hbar\mbox{ ps}^{-1})$ \\
\hline
\hline
Composition & 
         $^{+0.0216}_{-0.0131}$ & $^{+0.0225}_{-0.0131}$ & $^{+0.0295}_{-0.0310}$ \\
\hline
Other sources    & $\pm 0.0068$ & $\pm 0.0074$ & $\pm 0.0146$ \\
\hline
\hline
Total  & 
	   $^{+0.0226}_{-0.0147}$ & $^{+0.0237}_{-0.0150}$ & $^{+0.0329}_{-0.0343}$ \\ 
\hline
\end{tabular} 
\caption{ 
	The breakdown of the systematic uncertainties.
	\label{tab:systematics}
	}
\end{table}


\newpage

\end{document}